\documentclass{emulateapj}
\usepackage{amssymb, amsmath, epsfig, rotating}

\def\apj{ApJ}

\def\apjs{ApJS}
\def\aj{AJ}

\def\mnras{MNRAS}

\def\aap{{A\&A}}

\begin{document}

\title{A 189 MHz, 2400 Square Degree Polarization Survey with the Murchison Widefield Array 32-element Prototype}
\author{G. Bernardi\altaffilmark{1}, L.J. Greenhill\altaffilmark{1} , D.A. Mitchell\altaffilmark{2,3}, S.M. Ord\altaffilmark{4}, B.J. Hazelton\altaffilmark{5}, B.M. Gaensler\altaffilmark{6,3}, A. de Oliveira-CostaC\altaffilmark{1}, M.F. Morales\altaffilmark{5}, N. Udaya Shankar\altaffilmark{7}, R. Subrahmanyan\altaffilmark{3,7}, R.B. Wayth\altaffilmark{3,4}, E. Lenc\altaffilmark{6,3}, C.L. Williams\altaffilmark{8}, W. Arcus\altaffilmark{4}, S.B. Arora\altaffilmark{4}, D.G. Barnes\altaffilmark{9}, J.D. Bowman\altaffilmark{10}, F.H. Briggs\altaffilmark{11,3}, J.D. Bunton\altaffilmark{12}, R.J. Cappallo\altaffilmark{13}, B.E. Corey\altaffilmark{13}, A. Deshpande\altaffilmark{7}, L. deSouza\altaffilmark{12,6}, D. Emrich\altaffilmark{4}, R. Goeke\altaffilmark{8}, D. Herne\altaffilmark{4}, J.N. Hewitt\altaffilmark{8}, M. Johnston-Hollitt\altaffilmark{14}, D. Kaplan\altaffilmark{15}, J.C. Kasper\altaffilmark{1}, B.B. Kincaid\altaffilmark{13}, R. Koenig\altaffilmark{12}, E. Kratzenberg\altaffilmark{13}, C.J. Lonsdale\altaffilmark{13}, M.J. Lynch\altaffilmark{4}, S.R. McWhirter\altaffilmark{13}, E. Morgan\altaffilmark{8}, D. Oberoi\altaffilmark{18}, J. Pathikulangara\altaffilmark{12}, T. Prabu\altaffilmark{7}, R.A. Remillard\altaffilmark{8}, A.E.E. Rogers\altaffilmark{13}, A. Roshi\altaffilmark{7}, J.E. Salah\altaffilmark{13}, R.J. Sault\altaffilmark{2}, K.S. Srivani\altaffilmark{7}, J. Stevens\altaffilmark{16}, S.J. Tingay\altaffilmark{4,3}, M. Waterson\altaffilmark{11,4}, R.L. Webster\altaffilmark{2,3}, A.R. Whitney\altaffilmark{13}, A. Williams\altaffilmark{17} \& J.S.B. Wyithe\altaffilmark{2,3}}

\altaffiltext{1} {Harvard-Smithsonian Center for Astrophysics, Garden Street 60, Cambridge, MA, 02138; gbernardi@cfa.harvard.edu}
\altaffiltext{2} {The University of Melbourne, Melbourne, Australia}
\altaffiltext{3} {ARC Centre of Excellence for All-sky Astrophysics (CAASTRO)}
\altaffiltext{4} {International Centre for Radio Astronomy Research, Curtin University, Perth, Australia}
\altaffiltext{5} {University of Washington, Seattle, USA}
\altaffiltext{6} {Sydney Institute for Astronomy, School of Physics, The University of Sydney, Sydney, Australia}
\altaffiltext{7} {Raman Research Institute, Bangalore, India}
\altaffiltext{8} {MIT Kavli Institute for Astrophysics and Space Research, Cambridge, USA}
\altaffiltext{9} {Swinburne University of Technology, Melbourne, Australia}
\altaffiltext{10} {Arizona State University, Tempe, USA}
\altaffiltext{11} {Australian National University, Canberra, Australia}
\altaffiltext{12} {CSIRO Astronomy and Space Science, Australia}
\altaffiltext{13} {MIT Haystack Observatory, Westford, USA}
\altaffiltext{14} {Victoria University of Wellington, New Zealand}
\altaffiltext{15} {University of Wisconsin-Milwaukee, Milwaukee, USA}
\altaffiltext{16} {University of Tasmania, Hobart, Australia}
\altaffiltext{17} {University of Western Australia}
\altaffiltext{18} {National Centre for Radio Astrophysics, Tata Institute for Fundamental Research, Pune, INDIA.}

\begin{abstract} 
We present a Stokes $I$, $Q$ and $U$ survey at 189~MHz with the Murchison Widefield Array 32-element prototype covering 2400 square degrees. 
The survey has a 15.6~arcmin angular resolution and achieves a noise level of 15~mJy~beam$^{-1}$.
We demonstrate a novel interferometric data analysis that involves calibration of drift scan data, integration through the co-addition of warped snapshot images and deconvolution of the point spread function through forward modeling.
We present a point source catalogue down to a flux limit of 4~Jy. We detect polarization from only one of the sources, PMN~J0351-2744, at a level of $1.8 \pm 0.4$\%, whereas the remaining sources have a polarization fraction below 2\%.
Compared to a reported average value of 7\% at 1.4~GHz, the polarization fraction of compact sources significantly decreases at low frequencies.
We find a wealth of diffuse polarized emission across a large area of the survey with a maximum peak of $\sim$13~K, primarily with positive rotation measure values smaller than +10~rad~m$^{-2}$. The small values observed indicate that the emission is likely to have a local origin (closer than a few hundred parsecs). 
There is a large sky area at $\alpha \geq 2^{\rm h} 30^{\rm m}$ where the diffuse polarized emission rms is fainter than 1~K. 
Within this area of low Galactic polarization we characterize the foreground properties in a cold sky patch at $(\alpha, \delta) = (4^{\rm h}, -27^\circ.6)$ in terms of three dimensional power spectra.
\end{abstract}


\section{introduction}

The study of the Epoch of Reionization (EoR) through observations of the redshifted 21~cm Hydrogen line is motivating a renaissance in low frequency radio astronomy. 

New radio telescopes operating below 200~MHz have been built or are currently under construction in order to take advantage of this scientific opportunity: the Giant Metrewave Radio Telescope (GMRT\footnote{http://www.gmrt.ncra.tifr.res.in}), the Low Frequency Array (LOFAR\footnote{http://www.lofar.org}), the Murchison Widefield Array \citep[MWA,][]{lonsdale09,tingay12}, the Large aperture Experiment to detect the Dark Ages \citep[LEDA\footnote{http://www.cfa.harvard.edu/LEDA},][]{greenhill12} and the Precision Array to Probe the Epoch of Reionization \citep[PAPER,][]{parsons10}. The Square Kilometer Array (SKA\footnote{http://www.skatelescope.org}) and Hydrogen Epoch of Reionization Arrays (HERA\footnote{http://reionization.org}) will represent the future development of low frequency radio astronomy, built from the experience derived from the current instrumentation.

The low frequency radio sky has been surveyed since the 1950s in order to characterize the radio point source population. The 6C survey \citep[]{hales93} covered the whole northern sky above $\delta = +30^\circ$ with a $\sim$4~arcmin resolution and down to 200~mJy. The 7C survey extended the 6C catalogue to almost 1.7~sr sky coverage with $\sim$70~arcsec resolution \cite[]{hales07}. 
At the lowest frequency end, a VLA 74~MHz survey of the sky above $\delta = -30^\circ$ was carried out by \cite{cohen07}, achieving a 100~mJy sensitivity at 80~arcsec resolution. 

The southern hemisphere has been explored, for example,  by \cite{slee77} and \cite{slee95} who observed a sample of selected extragalactic sources at 80~MHz and 160~MHz. At 160~MHz, the source catalogue includes sources down to 2~Jy measured with a few arcmin resolution, but it claims completeness only down to 4~Jy. The Mauritius Radio Telescope survey covered $\sim$1~sr of the southern sky at 151~MHz down to a 300~mJy sensitivity at a few arcmin resolution \citep[]{nayak10}.

New dipole arrays have recently become operational and started to survey the low frequency radio sky. \cite{jacobs11} report PAPER measurements of compact sources at 145~MHz, down to 10~Jy with 26~arcmin resolution.  \cite{williams12} have recently carried out deep integrations of a field located at $\delta = -10^\circ$ with the MWA 32 element prototype.

None of the aforementioned studies investigated Galactic diffuse emission and polarization. In particular, Galactic and extragalactic polarized emission  is poorly known at frequencies below 200 MHz and their characteristics cannot be extrapolated directly from higher frequency measurements. Recent observations at 150~MHz \citep[][]{pen09,bernardi10} constrained the Galactic polarized emission to be below 1~K rms at high Galactic latitudes, weaker than expected from a direct extrapolation of the 350~MHz data. Their limited sky coverage prevents, however, from drawing more extensive conclusions about the prominence of polarized foregrounds. No data are available to assess the polarization properties of extragalactic radio sources below 200~MHz.

Upcoming observations of the redshifted 21~cm Hydrogen line from the EoR will require an unprecedented precision in mapping the low frequency sky in all its components, namely compact sources, diffuse total intensity and polarized Galactic emission. The greatest challenge in EoR observations is indeed the accurate subtraction of foreground sources which are two or three orders of magnitude brighter than the expected signal \citep[i.e.,][]{morales04,wang06,ali08,jelic08,bernardi09,bowman09,harker09,liu09,pen09,bernardi10,paciga11,ghosh12}, 
with bright compact radio galaxies which may need to be subtracted with a precision up to one part over ten thousand.

In this paper we present a large area sky survey with the MWA 32~element prototype in the 170-200~MHz range with the aim of improving the knoweldge of the foreground components for EoR observations and improving the global sky model for MWA calibration. 
We test novel calibration and imaging techniques for low frequency dipole arrays, investigate the properties of foregrounds and apply subtraction techniques relevant for EoR experiments. We focus particularly on Galactic polarized emission, providing constraints on its distribution as a function of Galactic latitude, and investigate the polarization properties of a sample of bright radio sources.

The paper is organized as follows: the observations and the data reduction are described in Section~\ref{obs_red}, the survey results in Section~\ref{survey_results} and the conclusions are presented in Section~\ref{conc}.

\section{Observations and data reduction}
\label{obs_red}

The observations were carried out with the MWA 32 element (``tile", 32T) prototype, located in the outback of Western Australia at the Murchison Radio Observatory, and took place on September 21st 2010, for a total of $\sim$8~hours during the night, starting at UT = 14$^{\rm h}$.

The 32 tiles were arranged in a circular configuration which provides fairly uniform $uv$ coverage up to the maximum baseline of $\sim$350~m. Each tile consists of 16 dual-polarization, active dipole antennas laid out over a metal mesh ground screen in a $4 \times 4$ grid with a 1.1 meter centre-to-centre spacing and can be steered electronically through an analogue beam former that introduces appropriate delays and attenuations for each individual dipole \citep[]{lonsdale09,tingay12}. 

The survey was taken in a ``drift-scan" mode, i.e. the tiles always pointed to zenith without any change of the beam former delays with time. Table~\ref{obs_table} summarizes the main characteristics of the survey. 
Three tiles were not functional during the observations and were therefore discarded.
The data were recorded as consecutive segments of 5~min integrations. We refer to each of these segments as a ``snapshot" throughout the paper.
\begin{table}	
\caption[]{Summary of the observational setup.} 
\label{obs_table}
\begin{tabular}{l l}        
\hline\hline 
Central declination			& -26$^\circ$ 45'\\
Right ascension coverage				& $21^{\rm h} < \alpha < 24^{\rm h}$ \& $0^{\rm h} < \alpha < 6^{\rm h}$\\
Central frequency  			& 189~MHz\\
Frequency resolution			& 40~kHz\\
Bandwidth				& 30.72~MHz\\
Tile field of view at the			& \\
Half Power Beam Width			& $20^\circ \times 20^\circ$\\
Time resolution				& 8~sec\\
Angular resolution 			& $15.6 \times 15.6 \sec(\delta + 26^\circ.7)$~arcmin\\
\hline
\end{tabular}
   \end{table}

\subsection{Primary beam, flux and bandpass calibration}
\label{primary_beam}

Calibration was carried out using the Real-Time calibration and imaging System \citep[RTS, ][]{mitchell08,ord10}, in an off-line mode. The source PMN~J0444-2809, observed towards the end of the drift scan, was used to set the absolute flux scale. Its flux is 45~Jy at 160~MHz with a spectral index of 0.81 \citep[]{slee77}, tied to the \cite{baars77} flux scale. The uncertainty on the absolute calibration is 5\%. 

Interferometric calibration involves solving for antenna-based complex gains as a function of time and frequency and using these gains to correct the data. The 32T array does not have sufficient sensitivity to solve for gains at the highest time and frequency resolution but the gains are expected to change slowly on these time scales, therefore we performed the calibration by considering the frequency and time gain responses as if they were decoupled.

Throughout the paper we will use the measurement equation formalism which describes the tile gain by $2\times2$ complex Jones matrices \citep[]{hamaker96,smirnov11}. 

We split the 30.72~MHz bandwidth into four 7.68~MHz bands and we used PMN~J0444-2809 to measure each of the four direction independent bandpass gains. Visibilities were rotated towards the source and, for the real and imaginary component of each gain element, a polynomial fit was performed for each of the four bands. 

After the bandpass was applied, we fitted for the complex Jones matrices. The bandpass solutions and complex Jones matrices derived according to this procedure were applied to the full visibility dataset. This correction does not account for time variations of the bandpass, however, we estimated them to be within 10\% based on bandpass fits performed on the source PMN~J2107-2526.

PMN~J0444-2809 was also used to constrain the tile beam through drift scan observations. One of the difficulties of calibrating low frequency dipole arrays is the precise knowledge of the primary beam when it is formed by a hierarchical cluster of single dipole elements, i.e. by the combination of the dipole beam and the beam former response. Although the primary beams of the MWA elements are reasonably equal to each other due to the same size and orientation of the tiles, differences can still arise from different delays attributed to each tile and errors in the delay lines of the analog beamformer. The drift-scan observing mode offers a way to disentangle the degeneracy between the sky brightness distribution and the variations of the tile beams, because they remain fixed with time. 

We described the tile beam as the numerical co-addition of individual dipoles - for which an analytic model can be used - and validated the model by computing calibration solutions towards PMN~J0444-2809 while it moved through the primary beam. We found an average agreement to within 2\% between the data and the model across the whole observing band.

\subsection{Imaging}
\label{sec_imaging}

The visibility data were first flagged for possible RFI contamination using a median filter \citep[]{mitchell10}. Less than 0.1\% of the data was discarded, confirming the excellent radio quiet environment of the Murchison Radio Observatory. After flagging, bandpass and Jones matrix corrections were applied and each snapshot was Fourier transformed into a $20^\circ$ wide image, i.e. the size of the half power beam width at the average frequency of 189~MHz. 

The visibility data were averaged over 0.64~MHz channels in order to prevent bandwidth smearing of sources at the field edge. The sky brightness distribution was resampled into the Healpix frame \citep[]{gorski05} and each 0.64~MHz snapshot image was weighted by the primary beam response. 
The snapshots are then co-added in the image domain and eventually converted to Stokes $I$, $Q$ and $U$ parameters to form the final dirty mosaic, which covers $\sim$2400 square degrees.
A Gaussian taper was applied to the visibilities to down weight baselines shorter than 40~wavelengths in order to suppress sidelobes from Galactic diffuse emission. 
Wide field polarization corrections are performed in the resampling stage \citep[]{ord10}. 

We quantified the calibration accuracy by measuring the leakage from Stokes $I$ into Stokes $Q$ and $U$ in an individual snapshot. We first peeled PMN~J0444-2809 from the visibility data. The word ``peeling" indicates the subtraction of the source model multiplied by its best fit Jones matrices, therefore subtracting the source contribution from all four instrumental polarizations. We looked for source peaks in the frequency averaged Stokes $Q$ and $U$ images corresponding to sources brighter than $\sim$3~Jy in total intensity. No source peak was found, from which we estimated the polarization leakage to be, on average, lower than 1.8\% across the field of view.

\subsection{Deconvolution}
\label{forward_modeling}

Several deconvolution methods for fixed dipole arrays have recently been developed to account for a point spread function (PSF) that is spatially dependent \citep[]{pindor11,bernardi11,sullivan12}. The PSF becomes spatially variable within each snapshot and from snapshot to snapshot when individual images are warped and resampled to a constant right ascension frame. All the total intensity sky emission in the dirty mosaic originates from point sources apart from the radio galaxy Fornax~A, which is located just outside the half-power beam width. Therefore we used the forward modeling technique of \cite{bernardi11} developed for point source subtraction/deconvolution.

According to their formalism, each point source can be modeled by a three parameter vector ${\bf x}$ containing its position  $(\alpha$,$\delta)$ and flux and the deconvolution can be performed by iteratively solving the following algebraic system of equations:
\begin{eqnarray}
\Delta {\bf x} = ({\bf J}^T {\bf W J})^{-1} {\bf J}^T { \bf W} \Delta {\bf m},
\label{eq_minimization}
\end{eqnarray}
where $\Delta {\bf x}$ is the vector of parameter estimates, ${\bf J}$ is, here, the Jacobian matrix which contains the derivatives of the forward model-synthesized beam with respect to the parameters computed at the current parameter estimate ${\bf x}$, ${\bf W}$ is the weight matrix and $\Delta {\bf m}$ is the difference between the data and the forward model. We note that a similar technique, without accounting for direction dependent effects, is implemented in the Newstar package \citep[]{noordam94}.

\cite{bernardi11} fitted sources simultaneously in order to minimize their sidelobe contribution, but here, since the dynamic range is modest, we followed a hierarchical approach where the brightest sources were fitted first individually and then subtracted from the visibility data when the fainter sources were fitted. 
We began by identifying sources brighter than 10~Jy and fitting their initial position and flux through forward modeling as described above. The eight scans where the source was closest to zenith were used for the fit. A residual mosaic was created by subtracting the best fit model from the visibility data. In creating the residual mosaic we performed a proper source peeling \citep[]{mitchell08} to correct for possible antenna gain variations on small time scales. 

The procedure of source identification and fit was repeated on the residual mosaic until the deconvolution was stopped at a threshold of 4~Jy.  
Sources outside the field of view were subtracted when they generated sidelobes running through the field of view.

All the sources subtracted were restored back by using a Gaussian beam of 15.6~arcmin. We note that, in this way, the deconvolution from the array response, the source fit and the subtraction all became part of a single step. 

\begin{figure*}
\epsfig{file=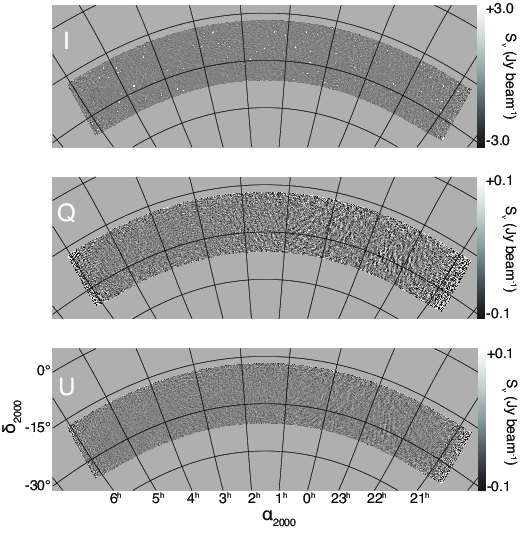,width=18cm}
\caption{Survey images in total intensity (top panel), Stokes $Q$ (middle panel) and Stokes $U$ (bottom panel) at the central frequency of 189~MHz. The total bandwidth used is 30.72~MHz. The Stokes $Q$ and $U$ images are shown at Faraday depth $\phi = 0$~rad~m$^{-2}$. A cartesian cylindrical coordinate projection is used. The image pixel size is 3.4~arcmin.\label{fig:final_mosaic}}
\end{figure*}
\begin{figure*}
\epsfig{file=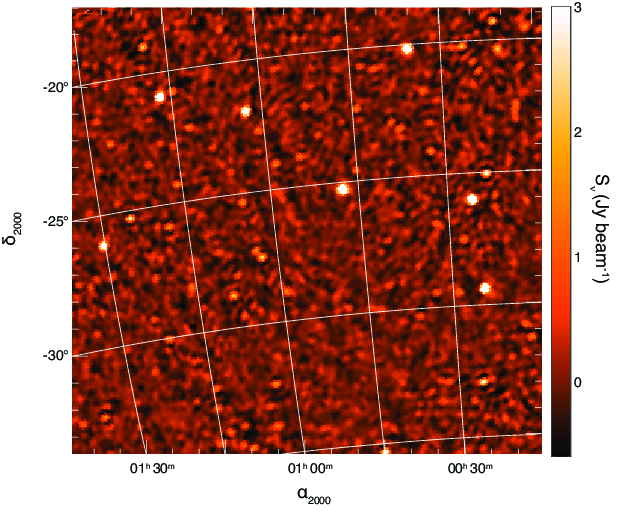,width=18cm}
\caption{A small region of the survey centred at $\alpha = 1^{\rm h}10^{\rm m}$. \label{fig:zoomin}}
\end{figure*}

\subsection{Rotation Measure synthesis}

Rotation Measure (RM) synthesis \citep[]{brentjens05} is a technique to measure polarized emission that takes advantage of the Fourier relationship between the polarized surface brightness $P(\lambda^2)$ and the polarized surface brightness per unit of Faraday depth $F(\phi)$ \citep[]{burn66,sokoloff98}:
\begin{eqnarray}
	P(\lambda^2) = W(\lambda^2) \int^{+\infty}_{-\infty} F(\phi) \, e^{-2i \phi \lambda^2} d \phi, \nonumber
\label{rm_synthesis_eq}
\end{eqnarray}
where $\lambda$ is the observing wavelength, $W(\lambda^2)$ is a weighting function and $\phi$ is the Faraday depth. The output of the RM synthesis is a cube of polarized images at selected values of Faraday depth. The Fourier transform of $W(\lambda^2)$ gives the RM spread function (RMSF), which determines the resolution in Faraday depth. The RMSF width depends uniquely upon the maximum $\lambda^2$ distance - analogous to the baseline length in imaging synthesis. With the present frequency coverage, the RMSF width is $\sim$4.3~rad~m$^{-2}$.

The channel width over which the visibility data are averaged, sets the sensitivity to the maximum RM, i.e. sources with higher RMs will suffer from bandwidth depolarization. We intially carried out a search for RM values as high as $| \phi_{max} | \sim 1200$~rad~m$^{-2}$ using the 40~kHz resolution without finding any high RM sources.
We therefore restricted the search to a cube which covers $-50 < \phi < 50$~rad~m$^{-2}$ in steps of 1~rad~m$^{-2}$ and this was used for the analysis presented in the following sections.

The RM cube was deconvolved according to the RMclean algorithm of \cite{heald09}.

\section{Results}
\label{survey_results}

\subsection{Bright source sample}
\label{culgoora_comparison}

The deconvolved survey mosaic is displayed in Figure~\ref{fig:final_mosaic}, with a small region of the image around $\alpha = 1^{\rm h}$ in Figure~\ref{fig:zoomin}. The distribution of pixel intensities of the residual mosaic (after sources brighter than 4~Jy were removed) was found to have an rms of $\sim$200~mJy~beam$^{-1}$, which can be considered the noise floor of the survey in total intensity. The sensitivity has a $\sim$50\% variation in declination because of the primary beam correction, with its maximum at $\delta = -26^\circ.7$, and changes with the number of snapshots contributing to each image pixel, therefore it is lower at the edges of the survey, i.e. for $\alpha > 5^{\rm h} 45^{\rm m}$ and $\alpha < 21^{\rm h}$.

Using the radio source counts from the 6C catalogue at 150~MHz \citep[]{hales93}, \cite{williams12} estimated the classical confusion noise for the MWA 32T array to be 160~mJy level at 154~MHz. Assuming a spectral index $\alpha = -0.7$ for the source population, a classical confusion noise at 189~MHz is expected to be $\sim$140~mJy, about 30\% lower than the 200~mJy rms level measured in the survey. The sensitivity of the Stokes $I$ image is therefore limited by confusion, defined to be the superposition of classical source confusion and noise due to the coupling of array sidelobes and the sky brightness distribution outside the field of view. Calibration errors may also contribute to the noise, but the level is difficult to quantify.

The frequency averaged Stokes $Q$ and $U$ images seen in Figure~\ref{fig:final_mosaic} appear mostly featureless apart from  diffuse structure at $23^{\rm h} < \alpha < 24^{\rm h}$. The distribution of their pixel intensities has a standard deviation of 15~mJy~beam$^{-1}$, higher than the expected $\sim$7~mJy~beam$^{-1}$ thermal noise. Throughout the paper we will adopt 15~mJy~beam$^{-1}$ as a conservative noise estimate.

We compiled a catalogue of 137 Stokes~I sources brighter than 4~Jy (Table~\ref{source_cat}). The source search was not carried out in a blind fashion, but started from the subtraction of the brightest sources down to the faintest ones (cf. Section~\ref{forward_modeling}). We limited our analysis to the bright sample of sources to avoid blending effects due to the limited resolution of the 32T array. All the sources were matched to the closest source of the PMN catalogue \citep[]{griffith93} within one beam size. The best fit source positions showed an average offset of $(\alpha,\delta) \sim (4,3)$~arcmin from the catalogue positions. The astrometry precision is limited by time dependent errors that were not corrected for as we selfcalibrated only on PMN~J0444-2809 and by ionospheric effects. \cite{williams12} found similar displacements in their analysis of the 32T data at the same frequencies.
\begin{table}	
\caption[]{Catalogue of sources brighter than 4~Jy at 189~MHz. The asterisk indicates the first measurements in the 100-200~MHz band. The errors are derived from the standard deviation of the distribution of pixel intensities in a $\sim$10~$\times$~10~arcmin area centred on each source after it was subtracted from the image.} 
\label{source_cat}
\begin{tabular}{l r l r}        
Source id & Flux (Jy) & Source id & Flux (Jy) \\
\hline\hline 
J2035-3454 & 18.5 $\pm$ 2.5 	& J0026-2004 &  6.0 $\pm$ 0.5\\
J2042-2855 &  4.8 $\pm$ 0.7 	& J0035-2004 & 11.9 $\pm$ 0.8\\
J2043-2633 &  7.1 $\pm$ 0.6 	& J0044-3530 &  7.8 $\pm$ 0.4\\
J2050-2948$^*$ &  4.3 $\pm$ 0.6 & J0047-2517 & 18.3 $\pm$ 1.1\\
J2051-2702$^*$ &  4.2 $\pm$ 0.3 & J0100-1749 &  6.1 $\pm$ 0.5\\
J2056-1956 & 13.8 $\pm$ 1.1 	& Cul 0100-221 & 11.4 $\pm$ 0.6\\
J2100-2828$^*$ &  4.6 $\pm$ 0.5 & J0102-2731 &  7.0 $\pm$ 0.4\\
J2101-1747 &  5.2 $\pm$ 0.6 	& J0108-2851$^*$ & 5.8 $\pm$ 0.3\\
J2101-2802 & 23.1 $\pm$ 1.5	& J0109-3447 &  4.4 $\pm$ 0.3\\
J2103-2749$^*$ &  5.7 $\pm$ 0.5	& J0116-2052 & 12.8 $\pm$ 0.7\\
J2107-1812$^*$ &  4.0 $\pm$ 0.5	& J0118-1849 &  5.1 $\pm$ 0.3\\
J2107-2526 & 47.7 $\pm$ 2.6	& J0118-2552 &  4.5 $\pm$ 0.3\\
J2110-3351 &  4.3 $\pm$ 0.5	& J0124-2517 &  7.1 $\pm$ 0.4\\
J2114-2541 &  7.0 $\pm$ 0.8	& J0130-2609 & 10.7 $\pm$ 0.8\\
J2114-3502 &  5.0 $\pm$ 0.4	& J0141-2706 &  8.9 $\pm$ 0.5\\
J2116-2055 & 17.4 $\pm$ 1.0	& J0150-2931 & 17.0 $\pm$ 1.0\\
J2118-3018 & 12.3 $\pm$ 0.7	& J0152-2940 &  4.9 $\pm$ 0.3\\
J2131-2036 &  8.6 $\pm$ 0.7	& J0156-3616$^*$ & 6.8 $\pm$ 0.4\\
J2131-3121 &  7.7 $\pm$ 0.5	& J0200-3053 & 18.7 $\pm$ 1.0\\
J2137-2042 & 12.5 $\pm$ 0.7	& J0237-1932 & 20.6 $\pm$ 1.1\\
J2138-1843 &  5.3 $\pm$ 0.4	& J0205-1801 &  5.3 $\pm$ 0.3\\
J2139-2556$^*$ & 5.3 $\pm$ 0.3	& J0211-2351 &  4.0 $\pm$ 0.4\\
J2151-1946 &  6.2 $\pm$ 0.8	& J0217-1757 &  4.6 $\pm$ 0.3\\
J2152-2828 &  6.7 $\pm$ 0.4	& J0218-2448 &  8.9 $\pm$ 0.5\\
J2155-3219$^*$ &  4.2 $\pm$ 0.3	& J0223-2819 &  8.3 $\pm$ 0.5\\
J2156-1813 & 12.1 $\pm$ 0.9	& J0225-2312 &  9.5 $\pm$ 0.6\\
J2206-1835 &  8.8 $\pm$ 0.7	& J0227-3037 &  4.7 $\pm$ 0.3\\
J2207-2003 &  5.6 $\pm$ 0.6	& J0231-2040 &  6.1 $\pm$ 0.4\\
J2208-3132$^*$ &  4.1 $\pm$ 0.3	& J0233-2321 &  9.5 $\pm$ 0.6\\
J2209-2331 &  4.7 $\pm$ 0.4	& Cul 0245-297 &  4.1 $\pm$ 0.4\\
J2214-2456$^*$ & 6.2 $\pm$ 0.5	& J0256-2324 & 13.1 $\pm$ 0.7\\
J2216-2803$^*$ & 5.5 $\pm$ 0.4	& J0258-2329$^*$ &  7.3 $\pm$ 0.9\\
J2218-3023$^*$ & 4.7 $\pm$ 0.3	& J0300-3413 &  8.0 $\pm$ 1.0\\
J2219-2756 & 11.5 $\pm$ 0.7	& J0307-2225 &  8.7 $\pm$ 0.6\\
J2237-1712 &  5.5 $\pm$ 1.0	& Cul 0313-271 &  8.2 $\pm$ 0.7\\
J2239-1720 &  4.9 $\pm$ 0.3	& J0328-2841 &  6.4 $\pm$ 0.5\\
J2245-1855 &  5.8 $\pm$ 0.4	& J0329-2600$^*$ &  6.9 $\pm$ 0.4\\
J2246-3044$^*$ &  4.8 $\pm$ 0.4	& J0338-3523$^*$ & 13.1 $\pm$ 1.0\\
J2250-2301 &  6.3 $\pm$ 0.6	& J0346-3422 & 18.5 $\pm$ 2.1\\
J2303-1841 &  4.1 $\pm$ 0.3	& J0351-2744 & 27.0 $\pm$ 1.5\\
J2304-3432 &  4.0 $\pm$ 0.3	& J0408-2418$^*$ &  7.3 $\pm$ 0.4\\
J2306-2507 &  5.6 $\pm$ 0.3	& J0409-1757 &  8.0 $\pm$ 0.5\\
J2310-2757 &  6.5 $\pm$ 0.4	& J0411-3513 &  4.4 $\pm$ 0.7\\
J2316-2729$^*$ & 5.0 $\pm$ 0.3	& J0413-3429 &  8.6 $\pm$ 0.6\\
J2319-2205 &  8.5 $\pm$ 0.5	& J0415-2929 &  9.2 $\pm$ 0.6\\
J2319-2727 & 13.4 $\pm$ 0.8	& J0416-2056 & 11.5 $\pm$ 0.6\\
J2320-1919$^*$ & 4.2 $\pm$ 0.4	& J0422-2616 &  5.4 $\pm$ 0.3\\
J2321-2410$^*$ & 6.0 $\pm$ 0.3	& J0426-2643 &  5.1 $\pm$ 0.4\\
J2324-2719 &  4.7 $\pm$ 0.3    	& J0423-3402 &  6.4 $\pm$ 0.4\\
J2328-2105 &  4.7 $\pm$ 0.6	& J0432-2956 &  5.7 $\pm$ 0.5\\
J2329-1923 &  6.1 $\pm$ 0.3	& J0437-2954 &  4.3 $\pm$ 0.4\\
J2329-2113 &  4.3 $\pm$ 0.6	& J0448-2032 &  5.4 $\pm$ 0.4\\
J2336-3444 &  6.6 $\pm$ 0.5	& J0452-2201$^*$ &  8.3 $\pm$ 0.7\\
J2341-3506 &  9.1 $\pm$ 0.6	& J0455-2034 & 17.7 $\pm$ 1.0\\
J2350-2457 & 10.5 $\pm$ 0.6	& J0455-3006 & 19.1 $\pm$ 1.1\\
J2356-3445 & 19.1 $\pm$ 0.9	& J0456-2159 & 10.6 $\pm$ 0.7\\
J0003-1727 &  8.6 $\pm$ 0.6	& J0458-3007 & 11.2 $\pm$ 0.7\\
J0003-3556 &  6.7 $\pm$ 0.4	& J0505-2826$^*$ &  8.0 $\pm$ 0.6\\
J0020-2014 &  4.6 $\pm$ 0.4	& J0505-2856$^*$ &  5.3 $\pm$ 0.3\\
J0021-1910 &  4.6 $\pm$ 0.4	& J0510-1838 & 14.4 $\pm$ 0.6\\
J0023-2502 &  8.2 $\pm$ 0.5	& J0511-2201 &  8.6 $\pm$ 0.6\\
J0024-2928 & 16.0 $\pm$ 0.9	& J0511-3315$^*$ &  6.5 $\pm$ 0.6\\
J0025-2602 & 19.0 $\pm$ 1.0	& J0513-3028 & 13.7 $\pm$ 0.9\\
J0025-3303 &  9.5 $\pm$ 0.7   	& J0521-2047 & 13.4 $\pm$ 0.8\\
\hline
\end{tabular}
\end{table}
\begin{table}	
\caption[]{Table~\ref{source_cat} continued} 
\begin{tabular}{l r l r}        
Source id & Flux (Jy) & Source id & Flux (Jy) \\
\hline\hline 
J0523-3251 &  8.0 $\pm$ 0.5     & J0543-2420 &  7.1 $\pm$ 0.4\\
J0539-3412$^*$ &  8.3 $\pm$ 0.6	& J0556-3222 &  8.4 $\pm$ 0.6\\
J0525-3242 &  5.5 $\pm$ 0.5    	& J0603-3144$^*$ &  8.2 $\pm$ 0.5\\
J0539-3412$^*$ &  8.3 $\pm$ 0.6	& J0603-3426 &  9.2 $\pm$ 0.7\\
J0540-3309$^*$ &  5.0 $\pm$ 0.5\\
\hline
\end{tabular}
\end{table}

\begin{figure}
{\epsfig{file=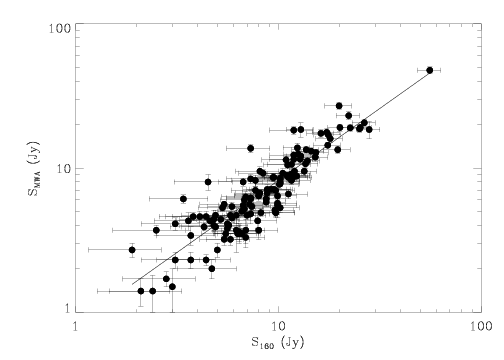, height=7.5cm, width=8.5cm}}
\caption{Comparison of the MWA 32T flux measurements at 189~MHz with the 160~MHz data \citep[]{slee77}. The solid line is the overall best fit spectral index $\alpha = -0.8$. The error bars do not include systematic errors. The absolute calibration error is 5\% at 189~MHz and 10\% at 160~MHz.\label{culgoora:comp}}
\end{figure}

Our flux measurements were compared with observations made with the Culgoora array \citep[]{slee77,slee95} at 160~MHz. Within the MWA 32T sky coverage we identified 136 common sources which provided a reference for the flux calibration of our survey (Figure~\ref{culgoora:comp}).

In order to compare ours and the 160~MHz flux measurements, we used an overall flux scaling given by $(188.8/160)^{\alpha}$, where $\alpha = -0.80 \pm 0.17$ is the overall best fit spectral index between the two frequencies. After the 189~MHz measurements were reported on the 160~MHz scale, we found an rms difference of $\sim 19$\% for sources brighter than 5~Jy at 189~MHz which decreases down to $\sim 16$\% for sources brighter than 10~Jy. The scatter between the two data sets broadens near to the 4~Jy threshold as a combination of the decrease in signal-to-noise ratio (SNR) and of the source selection criteria.

\subsection{Point source polarization}
\label{point_source_polarization}

Extragalactic radio galaxies exhibit an average polarization fraction of $\sim 7$\% at 1.4~GHz with peaks up to $\sim 20$\% \citep[]{taylor09}. 
Measurements of point source polarization are scarce at frequencies below $\sim$1~GHz. \cite{haverkorn03a} measured polarization for 15 sources brighter than 12~mJy at 350~MHz, with an average polarization fraction of $\sim$6\%. \cite{schnitzeler09} detected 23 polarized point sources brighter than 3~mJy at 350~MHz, with an average polarization fraction of $\sim 3$\%. Based on these limited numbers and their sky coverage, one would expect to have one polarized source every four square degrees with an average polarization fraction of a few percent. That would not seem to change significantly between 1.4~GHz and 350~MHz.

\begin{figure}
{\epsfig{file=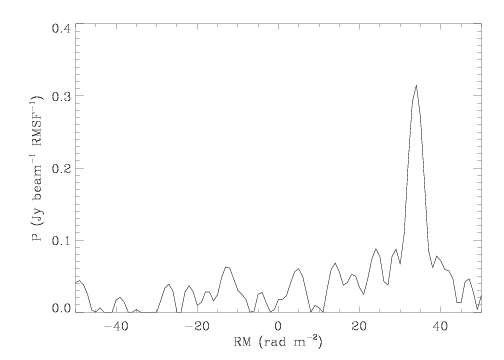, height=7.5cm, width=8.5cm}}
\caption{The Faraday depth spectrum of PMN~J0351-2744 at 189~MHz. The spectrum peaks at $\phi = +34 \pm 2$~rad~m$^{-2}$. The uncertainty is dominated by ionospheric Faraday rotation fluctuations.\label{fig:faraday_spectrum}}
\end{figure}

We used RM synthesis to investigate the point source polarization fraction at 189~MHz. We filtered out most of the large scale diffuse emission by retaining only the baselines longer than $\sim$40~wavelengths. The source PMN~J0351-2744 clearly showed a 320~mJy peak at $\phi \sim +34$~rad~m$^{-2}$  (see Figure~\ref{fig:faraday_spectrum}). This value is in agreement with the RM~$= +34.7 \pm 5.5$~rad~m$^{-2}$ measured by \cite{taylor09} at 1.4~GHz for this source whereas \cite{newton10} found an RM~$= -36.8 \pm 9$~rad~m$^{-2}$ at 1.4~GHz. The close agreement between the RM magnitude from the current work, \cite{taylor09} and \cite{newton10} suggests that there is a possible sign error in the determination of \cite{newton10}.

Observations of a linearly polarized source is essential for the correction of the unknown relative phase between the $p$ and $q$ polarizations \citep[]{sault96}. An uncalibrated phase between the two polarizations leads to a leakage of Stokes $U$ into $V$ and consequent depolarization. We used PMN~J0351-2744 to correct for the relative $p$-$q$ phase and obtain a full polarization calibration (the details are provided in Appendix~A).

\begin{figure}
{\epsfig{file=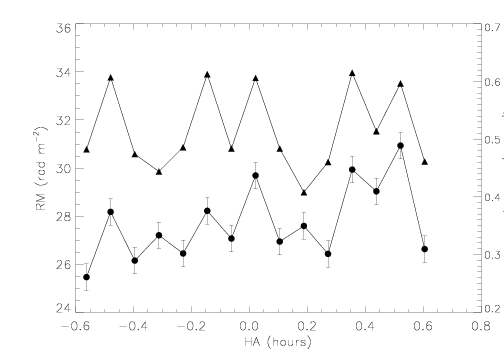, height=7.5cm, width=8.5cm}}
\caption{RM (triangles) and polarized intensity (circles) variations for PMN~J0351-2744 as a function of hour angle. \label{fig:ion_corr}}
\end{figure}

At 189~MHz, Faraday rotation due to the ionosphere can induce fluctuations in the measured RM, causing depolarization when the Stokes parameters are averaged over time. This effect is both time and direction dependent and, in order to be corrected across the very wide MWA field of view, would require a grid of calibration sources used to monitor the ionospheric behaviour.  In the present analysis we were limited to estimate the impact of ionospheric Faraday rotation towards PMN~J0351-2744 as it passed through the primary beam, by computing its RM for every 5~min snapshot (Figure~\ref{fig:ion_corr}). We found that the RM variations as a function of time have an rms of 1.8 rad~m$^{-2}$ for one hour of data. We note, however, the RM peak shows arcmin displacements as a function of time, suggesting that the observed RM variations might not only be due to ionospheric Faraday rotation. 

Temporal variations of the observed polarized intensity trace the RM variations fairly well. We estimated the depolarization due to RM fluctuations induced by the ionosphere by taking the ratio between the brightest peak at HA$\sim$0.5~h and the polarized intensity measured in the time integrated image and found a 53\% depolarization fraction. Correcting the time-integrated polarized intensity for this fraction, leads to a polarization fraction of $1.8 \pm 0.4$\% for PMN~J0351-2744.

We searched for polarization in all the sources brighter than 4~Jy listed in Table~\ref{source_cat} without any detection. Assuming five times the thermal noise as detection threshold, the polarization fraction is below $\sim$2\% for 4~Jy sources.
For the brightest sources the systematic error becomes a limiting factor over thermal sensitivity. We therefore adopted an average value of 2\% as upper limit on the polarization fraction for all the radio sources in our catalogue. 

Depolarization at low frequency is likely to be caused by both depth and beam depolarization. Depth depolarization (also called differential Faraday dispersion) occurs when the emitting and rotation regions are co-located in a ordered magnetic field. The polarization plane of the radiation emitted at the far side of the region undergoes a different amount of Faraday rotation compared to the polarized radiation coming from the near side, causing depolarization when the emission is integrated over the entire region. For a plasma with uniform density and uniform magnetic field, the polarization fraction $P'$ is \citep[]{burn66}:
\begin{eqnarray}
P' = p_0 \left| \frac{\sin{\phi \lambda^2}}{\phi \lambda^2} \right|
\label{depth_dep}
\end{eqnarray}
where $p_0$ is the intrinsic polarization fraction.

Beam depolarization occurs when the polarization angle changes significantly within the PSF. The change can be either intrinsic to the source or caused by an external foreground Faraday screen. Both magnetic field or electron density variations with cells smaller than the PSF produce depolarization according to \citep[]{sokoloff98}:
\begin{eqnarray}
P' = p_0 \, e^{-2 \sigma_{RM}^2 \lambda^4}
\end{eqnarray}
where $\sigma_{RM}$ is the RM dispersion across the source on the sky.

The depolarization ratio between 1.4~GHz and 189~MHz for PMN~J0351-2744 is $\frac{p_0}{P'} = 11$  and can be caused by either a $\phi = 1.2$~rad~m$^{-2}$ in the case of internal Faraday dispersion or by a $\sigma_{RM} \sim 0.4$~rad~m$^{-2}$ in case of beam depolarization.
\begin{figure}
\begin{center}
{\epsfig{file=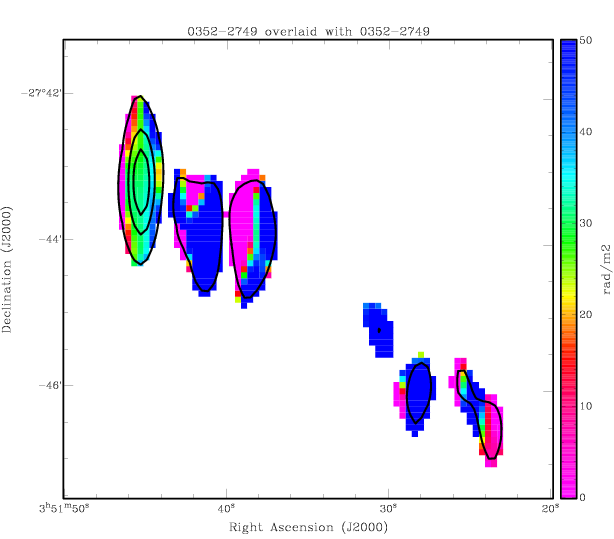,width=8.5cm, bbllx=0,bblly=5,bburx=613, bbury=500,clip=}}
\end{center}
\caption{RM map of PMN~J0351-2744 at 1.4~GHz observed with the Australia Telescope Compact Array. RM variations of several tens of rad~m$^{-2}$ occur within a couple of arcminutes.\label{fig:atca}}
\end{figure}

We re-analyzed archival data of PMN J0351-2744 observed with the Australia Telescope Compact Array at 1.4~GHz \citep[]{gaensler09} that resolve the source structure in RM and show variations across the source itself, spanning the $+10 < {\rm RM} < +50$~rad~m$^{-2}$ range (Figure~\ref{fig:atca}). This result further confirms that our polarization calibration gives the correct sign for the source RM at 189~MHz. 

The observed RM variations across the source have a $\sigma_{\rm RM} = 54$~rad~m$^{-2}$, sufficient to depolarize the source when integrated over the PSF of the 32T array, therefore beam depolarization seems a more likely explanation over depth depolarization in this case.

The polarization fraction of the remaining compact sources is below 2\%, whereas it is 7\% at 1.4~GHz. Their general depolarization mechanism is not well constrained by our observations. \cite{leahy87} estimated the Galactic RM contribution on 1-2~arcmin scales to a sample of selected 3C sources. The sample at $b > 50^\circ$ has $\sigma_{\rm RM} = 7$~rad~m$^{-2}$ and he found that no Galactic foreground could explain these fluctuations, whereas sources at $|b| < 10^\circ$ has $\sigma_{\rm RM} = 15$~rad~m$^{-2}$. He concluded that RM fluctuations at high Galactic latitudes happen at the source and, as we have shown above, they can beam depolarize the emission. 
On the other hand, internal Faraday dispersion cannot be ruled out either, because it can generate complete depolarization for $\phi > 1.2$~rad~m$^{-2}$ which is not uncommon in radio sources \cite[i.e.][]{osullivan13}. Multifrequency, higher resolution observations are needed to distinguish between the two mechanisms.

\subsection{Fornax A}
\label{fornax_a_section}

Fornax~A (NGC~1316) is a very well studied radio source, which lies at the outskirts of the Fornax cluster. At optical wavelengths it appears as a D-type galaxy \citep[]{schweizer80} and it has been been extensively observed at many radio frequencies. \cite{cameron71} and \cite{ekers83} resolved its structure  at 408 and 1415~MHz respectively, with arcminute resolution. They identified two radio lobes, a bridge that connects them and a compact core. \cite{fomalont89} studied its total intensity and polarized emission at 1.51~GHz with 14~arcsec angular resolution.

Fornax~A transits 11$^\circ$ away from zenith at the MWA location and our primary beam model is still valid at that distance. It was deconvolved by identifying CLEAN-like components at the center of each image pixel. A synthesized beam at the location of the pixel centre was generated for each component, a scaled version of the synthesized beam was subtracted and a new component searched for. This deconvolution scheme is therefore similar to the forward modeling described in Section~\ref{forward_modeling} if we exclude the fit for the position.
A residual rms floor of $\sim$2.5~Jy~beam$^{-1}$ was reached even without including correlation among adjacent pixels \citep[]{bhatnagar04}.

\begin{figure}
\begin{center}
{\epsfig{file=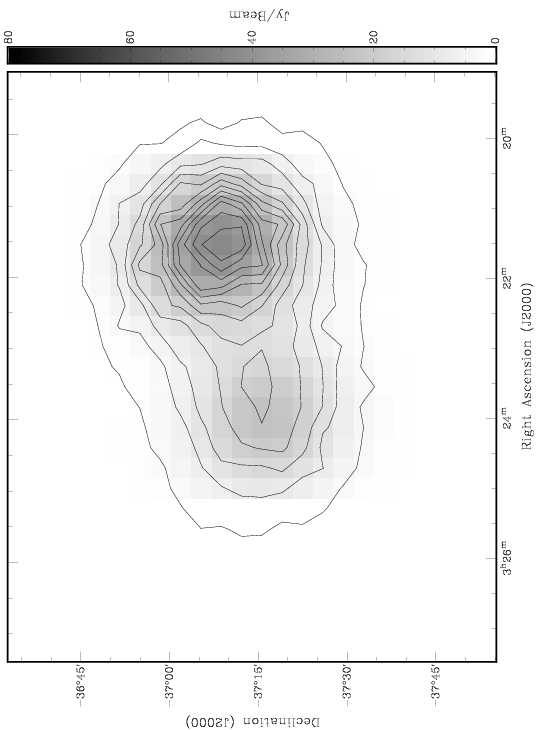, height=8.5cm, width=6.5cm, angle = 270}}
\end{center}
\caption{Fornax~A at 189~MHz (contours) overlaid on the VLA 1.51~GHz image \citep[]{fomalont89}. The VLA image was smoothed to the MWA 32T resolution of 15.6~arcmin. Contours are drawn between 10 and 220~Jy~beam$^{-1}$ in steps of 17.6~Jy~beam$^{-1}$. \label{fig:fornaxa}}
\end{figure}

Figure~\ref{fig:fornaxa} shows the 189~MHz contour map of Fornax A overlaid with the 1.51~GHz image \citep[]{fomalont89}, convolved to the MWA 32T resolution. The two images were overlapped by finding the best fit overall offset in $(\alpha,\delta)$ using the MIRIAD task imdiff \citep[]{sault95}. We corrected for an overall 4~arcmin offset in declination, compatible with the coordinate offset found in Section~\ref{culgoora_comparison} for point sources. 

A similar morphology emerges from the comparison of the two images. At 14~arcsec resolution, the radio lobes have sharp edges, with the west lobe brighter than the east lobe and an unresolved core emission. When it is smoothed down to the 32T resolution, only the two lobes remain visible and their sharp edges are smoothed out.

The size of the lobes at 189~MHz and 1.51~GHz is consistent, suggesting that little evolution occurs in the electron population. We notice that the peak of the faintest lobe is displaced by $\sim$7~arcmin between the two frequencies. This might be an indication of frequency evolution of the relativistic particles, but it is difficult to draw a firmer conclusion because of the limited angular resolution.

\begin{figure}
\begin{center}
{\epsfig{file=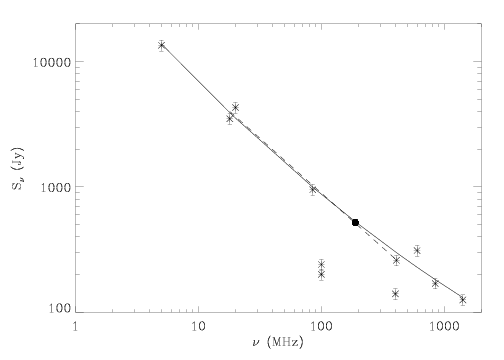, height=7.5cm, width=8.5cm}}
\end{center}
\caption{The frequency spectrum of Fornax~A between 5 and 1415~MHz. Asterisks are measurements taken from the literature at 5 \citep[]{ellis66}, 18 \citep[]{shain54}, 20 \citep[]{shain58}, 85 \citep[]{mills60}, 100 \citep[]{stanley50,bolton54}, 400 \citep[]{mcgee55}, 408 \cite[]{robertson73}, 600 \cite[]{piddington56}, 843 \cite[]{jones92} and 1415~MHz \citep[]{ekers83}. The filled circle is from this work. The solid line represents a second order polynomial fit in the 5-1400~MHz range. The dashed line represents a power law fit in the 30-400~MHz range, which encompasses the MWA frequency coverage (see text for details). \label{fig:fornaxa_spectrum}}
\end{figure}

We measured the integrated flux over the source to be $S_{189} = 519 \pm 26$~Jy and compared it with previous measurements at other frequencies (Figure~\ref{fig:fornaxa_spectrum}). 
Our measurement is in very good agreement with the spectrum between 5 and 1415~MHz. It appears very smooth and does not show indications of a turnover even at very low frequencies. Two data points at 100~MHz and one data point at 400~MHz appear unexpectedly low and might be affected by systematic errors. After excluding these values, a second order polynomial of the form
\begin{eqnarray}
\ln S_\nu = \sum_{i=0}^2 a_i (\ln \nu)^i
\end{eqnarray}
is the best fit to the frequency spectrum over three decades in frequency. If we included data between 30 and 400~MHz and fit a power law spectrum, we found a spectral index $\alpha_{30-400} = -0.88 \pm 0.05$. The rms difference between a power law and the best fit model is 0.3\% over the MWA frequency range. This is an important result because radio sources are going to be used to obtain beam calibration for low frequency arrays (cf. Section~\ref{primary_beam}) and most of them show deviations from a pure power law behaviour over several decades in frequency \citep[i.e, ][]{scaife12}. If they could be approximated with power law spectra with negligible errors, the requirements of accurate spectral models would be alleviated even when very precise beam calibration is required as for EoR observations. 

Finally, we investigated polarization from Fornax~A through RM synthesis. We found no evidence of polarized emission above three times the noise at any Faraday depth, indicating that the polarization fraction of the lobes is below 1\%.

At 1.51~GHz the lobes are 20\% polarized on average and show a spatially rich pattern with filaments and depolarization areas \citep[]{fomalont89}. RM values between -15 and +10~rad~m$^{-2}$ are common across the lobes with variations greater than 20~rad~m$^{-2}$ in the most depolarized regions \citep[]{fomalont89}. A conservative value of $\sigma_{\rm RM} = 5$~rad~m$^{-2}$ across the lobes would be sufficient to completely depolarize the emission at 189~MHz within the MWA 32T PSF, indicating that beam depolarization causes the observed depolarization at low frequencies.

\subsection{Diffuse polarization}
\label{wi_fi_pol}

Galactic polarized emission is a very powerful probe of the interstellar medium (ISM) and a possible contamination for EoR measurements \citep[]{jelic10,geil11,moore13}.

Diffuse polarized emission has been observed in large areas of the sky at 350~MHz on scales greater than 5~arcmin and at the level of several mK rms \citep[]{wieringa91,haverkorn03a,haverkorn03b}. A characteristic feature of low frequency radio polarization is the lack of correlation with total intensity emission, generally interpreted as the effect of foreground ISM clouds which rotate a smooth polarized background but leave the total intensity background untouched and, therefore, resolved out by the interferometric sampling \citep[]{wieringa91,gaensler01,bernardi03a}. 
Recent observations at 150~MHz revealed polarization rich \citep[]{bernardi09,bernardi10} as well as unpolarized regions \citep[]{pen09}. Polarized emission was found to be 
fainter than expected on the basis of a simple power law extrapolation from the 350~MHz data, suggesting that depolarization occurs. The current understanding of the global properties of the polarized ISM at low frequencies is, however, still limited by the scarcely available data.

Our observations cover a $20^\circ$ strip at Galactic latitude ($b < -20^\circ$), the largest polarization survey to date below 200~MHz. RM synthesis was extended to the whole survey in order to map Galactic polarization. All the compact sources brighter than 5~Jy in total intensity were peeled and, in this way, their instrumental polarization contribution to the RM synthesis was removed. The baselines shorter than 15 wavelengths were down weighted in order to suppress the very large scale emission. This weight gives sensitivity to emission up to $\sim$4$^\circ$~scales. 

Direction dependent instrumental polarization over the wide field of view is dealt with as described in \cite{ord10} and, assuming that the measured phase between the $p$ and $q$ polarizations is direction independent (cf. Section~\ref{point_source_polarization} and Appendix A), the calibration derived from PMN~J0351-2744 holds across the whole survey. 
The final Stokes $Q$ and $U$ maps were not corrected for ionospheric Faraday rotation, therefore they measure only a relative polarization angle.

\begin{figure*}
{\epsfig{file=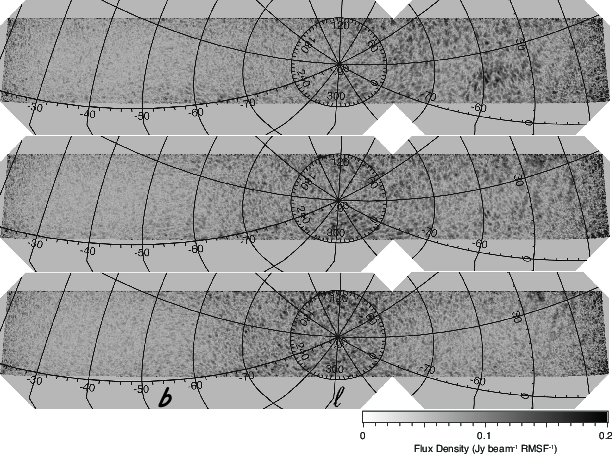, width=18cm}}
\caption{Polarized intensity images at Faraday depths $\phi$ = 0, +2 and +4~rad~m$^{-2}$ at the top, middle and bottom panels respectively. The representation is in Galactic coordinates and uses a slant orthographic projection \citep[]{calabretta02}. The maximum pixel value is 0.2~Jy~beam$^{-1}$~RMSF$^{-1}$ and the minimum pixel value is zero. The conversion factor is 1~Jy~beam$^{-1}$~RMSF$^{-1}$ = 44.4~K~RMSF$^{-1}$. Because the Stokes $Q$ and $U$ point source contribution was subtracted, we do not expect to see polarized point sources at $\phi = 0$~rad~m$^{-2}$, which is normally dominated by the instrumental polarization. If point sources are sufficiently polarized they will appear at their RM value (see Figure~\ref{fig:rm_1}, bottom panel). The bright structure centred at $(l,b) \sim (20^\circ,-60^\circ)$ seen in the top panel is an example of filamentary features with no total intensity counterpart.\label{fig:rm_0}}
\end{figure*}

\begin{figure*}
{\epsfig{file=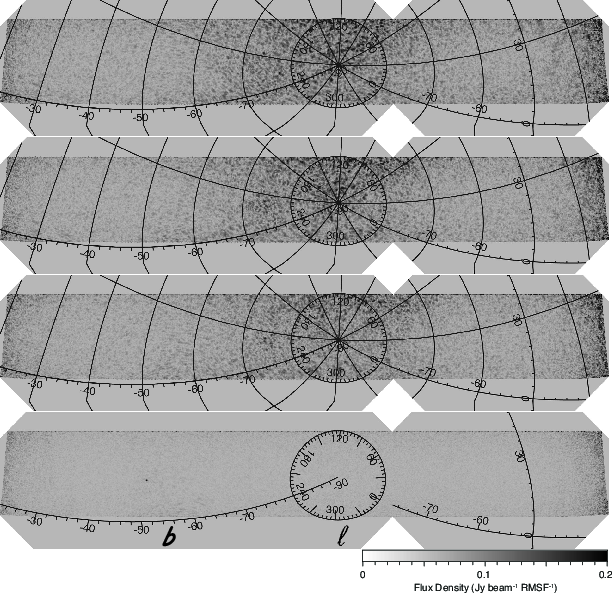, width=18cm}}
\caption{As Figure~\ref{fig:rm_0} but for Faraday depths $\phi$ = +6, +8, +10 and +34~rad~m$^{-2}$ from top to bottom panels respectively. Significant polarization structure is seen around the south Galactic pole. The bottom panel shows the polarized intensity from PMN~J0351-2744 at $(\ell,b) ~ (224^\circ:35',-50^\circ:15')$, in an otherwise empty Faraday depth slice. \label{fig:rm_1}}
\end{figure*}

We detected diffuse polarized emission over many degrees on the sky, varying with Galactic latitude over a factor of 20, from the brightest peaks at $\sim$13~K~RMSF$^{-1}$ down to the thermal noise (Figure~\ref{fig:rm_0} and \ref{fig:rm_1}). 

Interestingly, most of the emission occurs at a small range of Faraday depths across the whole survey, essentially in the $0 < \phi < +10$~rad~m$^{-2}$ range. It has a patchy and partly filamentary structure where the size of the patches can be as big as a couple of degrees down to the PSF size. Fainter, patchy polarized emission with peaks up to $\sim$2~K~RMSF$^{-1}$ appears at $\phi$ up to $\sim |20|$~rad~m$^{-2}$. The emission is interspersed by voids in polarization, similar to the canals described by \cite{haverkorn04b}. Such morphology was recently explained as sharp gradients in Stokes $Q$ and $U$ caused by localized high values of the gas density and magnetic field, resulting from vorticity or shear in the ISM in a sub-sonic regime \citep{gaensler11}. 

A detailed model of the ISM components is beyond the scope of the present paper and will be explored in a future publication, but we briefly consider the nature and the distance of the observed polarized emission.

The only polarization data available in the Southern Hemisphere at frequencies close to 189~MHz is the 408~MHz survey carried out with the Parkes telescope at $\sim$40~arcmin resolution \citep[]{mathewson65}. Its area overlaps with a large part of our survey, at $0^{\rm h} < \alpha < 3^{\rm h} 15^{\rm m}$ and $21^{\rm h} < \alpha < 24^{\rm h}$. Galactic polarization appears smooth at 408~MHz, with 1-3~K brightness levels over the area at $0^\circ < \ell < 30^\circ$ and $-80^\circ < b < -40^\circ$, extending to the area at $200^\circ < \ell < 215^\circ$ and $-80^\circ < b < -70^\circ$. Very little polarization is detected at $215^\circ < \ell < 230^\circ$ and $-70^\circ < b < -10^\circ$, whereas the peak of the emission is $\sim$6~K at $\sim 2^\circ$-wide area centred at ${\rm (\ell,b)} \sim (55^\circ, \, -70^\circ)$.

The comparison of polarization observations at different frequencies with different sampling in angular scales is complicated because it can mix beam and depth depolarization, however, both surveys show a similar trend. The small scale details are rather different though, for instance, at 189~MHz there is no correlation with the 6~K peak observed at 408~MHz. If we assume an average level of emission of $\sim$3~K at 408~MHz and the polarized emission to have a spectral index of $\beta = -2.6$ like the total intensity \citep[]{rogers08}, we would expect an average brightness of $\sim$20~K at 189~MHz, which is even higher than the brightest peaks observed. 

Depolarization at 189~MHz could be due to a resolved large-scale polarized background, depth and beam depolarization. The presence of depolarization indicates that polarized emission at 189~MHz comes from the foreground ISM compared to the 408~MHz data and has a more local origin.

\cite{landecker01} introduced the concept of ``polarization horizon" as the distance beyond which most of the emission is depolarized when it reaches the observer. Generally, the distance of the polarization horizon can be affected by both beam and depth depolarization, and both effects depend upon the observing frequency and the angular resolution. At 1.4~GHz the polarization horizon is $\sim1-3$~kpc \citep[]{bernardi03b,gaensler01,bernardi04} but it shrinks down to $\sim 600$~pc at 408~MHz \citep[]{brouw76} and at 350~MHz \citep[]{haverkorn04a}. 

We used the 189~MHz emission to constrain the distance of the polarization horizon at very low frequencies. Assuming a magnetic field of 2-4$~\mu$G and an electron density of 0.05-0.08~cm$^{-3}$ \citep[]{reynolds91}, the maximum ${\rm RM} = +10$~rad~m$^{-2}$ observed in our data constrains the polarized emission to be more local than $\sim$120~pc.

An independent estimate of the distance of the polarization horizon can be inferred from a comparison with RMs of pulsars with known distances. Given the high Galactic latitude coverage of the 189~MHz survey, there are only six pulsars with measured RM within the survey coverage \citep[]{han06,han99}. Four pulsars have RMs between +13 and +50~rad~m$^{-2}$ and a distance greater than 1.3~kpc and two of them are closer than few hundred parsecs and have RMs of +5 and +30~rad~m$^{-2}$ respectively. Given the very small sample available, we cannot draw statistically robust conclusions, but pulsar RMs have the same sign as the diffuse polarization and seem to support a distance of the polarization horizon in agreement with what was estimated by using RMs of the diffuse emission.

\subsection{Foreground characterization in a potential EoR window}

One of the main science drivers of low frequency radio observations is the measurement of the 21~cm line from the EoR. All the current models predict a cosmological signal that peaks at $\sim$10~mK in the 100-200~MHz \citep[i.e.][]{mcquinn06,mellema06}, therefore the deconvolution and subtraction of the bright foreground sky is the crucial step to measure the underlying faint cosmological signal \citep[for a review on the topic, see ][]{furlanetto06,morales10}. 

Section~\ref{culgoora_comparison} showed that forward modeling recovers the flux of bright sources fairly well, however, for precision spectral studies such as the EoR, the deconvolution should also not introduce any artifacts into the frequency domain. This requirement is difficult given the strong chromatic features of the MWA antennas. In this section we use some of the techniques developed for EoR power spectrum analyses to test the spectral properties of the bright source deconvolution and to characterize the remaining residuals. The image cubes we used in this section have not been corrected by the baseline distribution as they would need to in order to be compared with the EoR signal; we leave the full power spectrum analysis of the survey data with baseline weights and error propagation for a future paper.

The first step in the power spectrum analysis is to create a three dimensional cube in Fourier (or ${\bf k}$) space by mapping the frequency dimension to line-of-sight distance (i.e. the distance of the redshifted 21~cm line at that frequency) and then taking a three dimensional Fourier transform. We used the following definitions \citep[]{morales04}:

\begin{eqnarray}
k_\perp & = & \frac{2 \pi u}{D_M(z)}\\
k_\parallel & = & \frac{2 \pi H_\circ f_{HI}}{c \, (1+z)^2 \sqrt{\Omega_M (1+z)^3 + \Omega_\lambda}}
\label{k_def}
\end{eqnarray}
where $u$ is the baseline length in wavelengths. We have assumed a flat Universe $\Omega_k = 0$ and the transverse comoving distance $D_M(z)$ as:
\begin{eqnarray}
D_M(z) = \frac{(1+z) \, c}{H_\circ} \int_0^z \frac{d z'}{\sqrt{\Omega_M (1+z')^3 + \Omega_\lambda}}
\label{dm_def}
\end{eqnarray}
In Equation~\ref{k_def} and~\ref{dm_def}, $H_\circ$ is the Hubble constant, $f_{HI} = 1421$~MHz is the rest frequency of the 21~cm line, $\Omega_M$ and $\Omega_\lambda$ are the matter and dark energy content respectively.

\begin{figure*}
{\epsfig{file=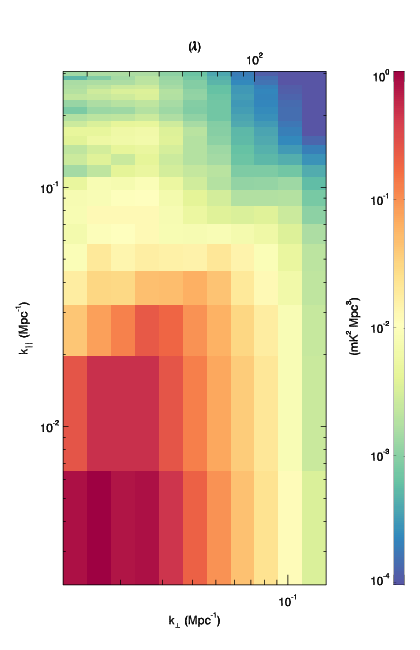, width=8.5cm}}
{\epsfig{file=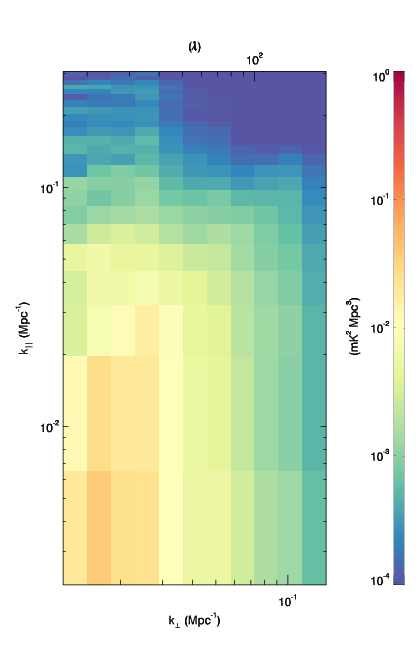, width=8.5cm}}
\caption{2D power spectra from the patch centred at $\alpha = 4^{\rm h}$ before (left) and after point source deconvolution and subtraction (right). The limits on the $k_{\perp} \times k_{\parallel}$ plots depend on the baseline coverage (baseline length in wavelength is indicated on the top axis) and the frequency resolution and bandwidth (30.72~MHz bandwidth with channel width $\Delta \nu = 0.64$~MHz gives a $k_\parallel$ range of $\sim 0.012 \textendash 0.63$~Mpc$^{-1}$). The three dimensional power cubes were averaged in logarithmic $k_\perp = \sqrt{k_x^2 + k_y^2}$ bins while the $k_\parallel$ bins were unchanged to allow the linear bin edges to be identified. The color scale in both spectra has been normalized to the peak of the power before deconvolution. 
\label{fig:coldpatch}}
\end{figure*}

While the EoR signal is expected to be spherically symmetric in $k$ space, the astrophysical sources have a very different geometry because they are expected to be spectrally smooth \citep[]{zaldarriaga04, morales04}. The difference in these geometries is most easily seen by averaging in circular shells in the $k_x$ and $k_y$ directions to produce two dimensional plots of power in $k_{\perp}$~vs~$k_{\parallel}$ space (Figure~\ref{fig:coldpatch}). These plots are a useful representation because the spectrally smooth foregrounds appear predominantly in the lowest $k_\parallel$ bin. A fraction of the observed power is scattered up by the chromatic response of the instrument into higher bins in a characteristic ``wedge" \citep[]{datta10,vedantham12,morales12,trott12}. This wedge leaves an ``EoR window" open that is relatively free of contamination. 

We computed the power spectrum from a $20^\circ \times 20^\circ$ sky patch centred at $\alpha = 4^{\rm h}$, which was previously indicated by \cite{bowman09} as an area with low Galactic foreground emission. This patch has an rms of 11.5~K in the frequency averaged image (Figure~\ref{fig:final_mosaic}). 

2D power spectra with 0.64~MHz frequency resolution are shown in Figure~\ref{fig:coldpatch} before and after the deconvolution of bright point sources, where the ratio between the Stokes $I$ and $Q$ rms per frequency channels is approximately a factor of five.

\begin{figure*}
{\epsfig{file=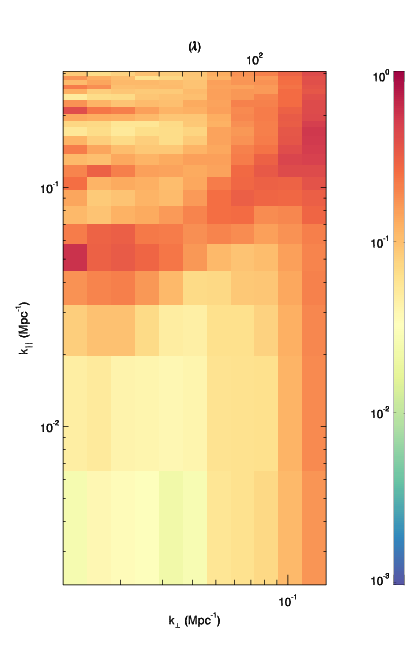, width=8.5cm}}
{\epsfig{file=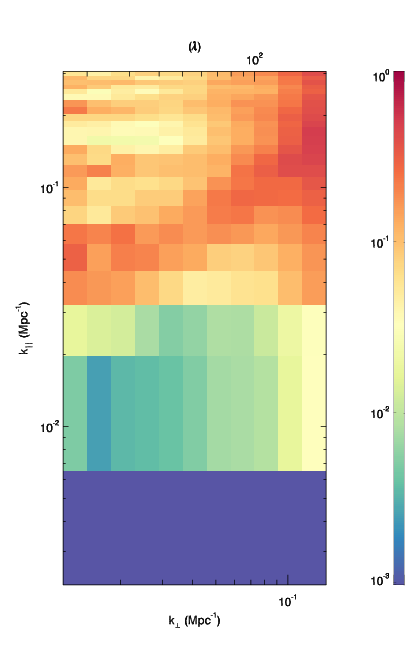, width=8.5cm}}
\caption{Ratio between 2D power spectra after the two foreground subtraction steps: the ratio between the power spectra before and after point source subtraction (left panel) and before and after filtering of the diffuse component (right panel). The orange diagonal stripe that emerges in both panels between $0.05 < k_\parallel < 0.2$~Mpc$^{-1}$ separates the EoR window from the area below, where most of the foregrounds are located.}
\label{fig:coldpatch_ratio}
\end{figure*}

The decrease in power at large $k_\perp$ is due to the drop off in the number of baselines at those scales, since the cubes have not been weighted to remove this effect. Point source deconvolution removed most of the power at small $k$ values, below the wedge (Figure~\ref{fig:coldpatch_ratio}). The EoR window above the wedge remained unaffected.

The residuals after compact source deconvolution are expected to be dominated by confusion-level astrophysical sources and diffuse emission, both with fairly smooth spectra.
To investigate the spectral properties of these residuals, we used a principal component analysis (PCA) on the deconvolved image cube, following the approach of \cite{deoliveiracosta08} and \cite{liu12}. The method determines the eigenvalues and eigenvectors associated with the frequency correlation matrix $C$ of the observed signal. If the observed pixel temperatures at each frequency $i$ are grouped in a vector $\bf x$, the correlation matrix $\bf{C}$ is defined as:
\begin{eqnarray}
{\bf C} = \frac{1}{N_{pix}} \sum_{n = 1}^{N_{pix}}{\bf x}_n {\bf x}_n^T 
\end{eqnarray}
where $N_{pix}$ is the number of pixels in each image.
If $N_f$ frequencies are observed, $\bf{C}$ is a $N_f \times N_f$ matrix. The PCA  determines the eigenvalues and eigenvectors of $\bf{C}$:
\begin{eqnarray}
{\bf C} = {\bf P}{\bf \Lambda}{\bf P}^T 
\end{eqnarray}
where the eigenvectors are the columns of ${\bf P}$ and ${\bf \Lambda}$ is a diagonal matrix containing the eigenvalues in decreasing order, i.e. $\Lambda_{ij} = \delta_{ij} \lambda_i$. 
The sky model at each frequency is represented by a vector ${\bf m}_i$ defined as:
\begin{eqnarray}
{\bf m}_i = {\bf P}{\bf P}^T {\bf x}_i.
\end{eqnarray}
The eigenvectors are sorted by the corresponding eigenvalues in decreasing order, such that the first few eigenvectors are the dominant spectral shapes in the image cubes and represent most of the power. We found that the first five eigenvectors account for 60\% of the power, indicating that the power in the image cubes is not as highly concentrated in just a few eigenvectors as it is in \cite{deoliveiracosta08} model, where the first three components contain $\sim$99\% of the total power. This may be caused by structure introduced by sidelobes not suppressed by the limited $uv$ coverage of the drift scan survey or imperfect calibration of the array frequency response.

\begin{figure}
{\epsfig{file=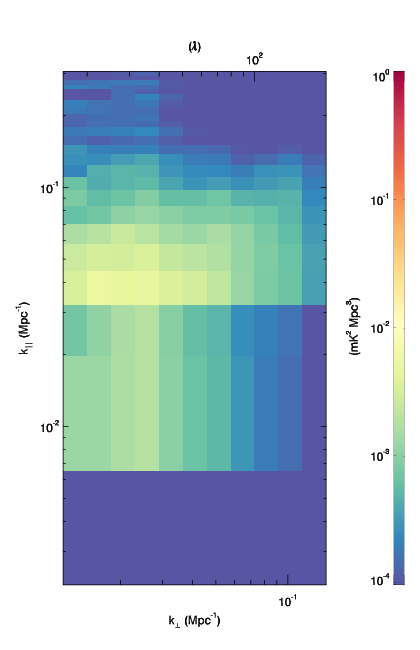, width=8.5cm}}
\caption{2D power spectrum for the image cube after deconvolution and subtraction of the first five principal components. The normalization is the same as Figure~\ref{fig:coldpatch}. The subtracted principal components primarily remove power in the first few $k_\parallel$ modes, indicating that the residuals after deconvolution are fairly smooth in frequency.}
\label{fig:pca_5comp}
\end{figure}

To evaluate the effect of the dominant modes on the power spectra, we subtracted the contribution of the first five eigenvectors from the residual image cube. The residual image at each frequency therefore becomes:
\begin{eqnarray}
{\bf x}_i - {\bf m}'_i = {\bf x}_i - {\bf P}'{\bf P}'^T {\bf x}_i	
\end{eqnarray}
where ${\bf P}'$ is an $5 \times N_f$ matrix which contains the first five eigenvectors. We plot the resulting power spectrum in Figure~\ref{fig:pca_5comp}. Removing the dominant spectral modes primarily removed power in the lowest few $k_\parallel$ bins, as expected for residuals that are dominated by smooth astrophysical sources (Figure~\ref{fig:pca_5comp} and Figure\ref{fig:coldpatch_ratio}). 

In Section~\ref{wi_fi_pol} we showed that the diffuse polarized emission decreases significantly at $\alpha \geq 2^{\rm h} 30^{\rm m}$. Polarized peaks up to 2~K~RMSF$^{-1}$ can be observed along individual lines of sight, but the polarized fluctuation rms decreases by approximately a factor of four when moving away from the south Galactic pole. 
We integrated the polarized emission over the $0 < \phi < +10$~rad~m$^{-2}$ range and found that the rms of polarization fluctuations is below 1~K. 

\cite{geil11} described a way to remove polarized emission that leaks into the total intensity and, potentially, corrupts the EoR signal. Polarized emission at small $\phi$ values can directly be filtered out by applying a high-pass filter in $\phi$ space and polarized emission at high $\phi$ values deconvolved from the RMSF in $\phi$ space. 
 
The characteristics of the polarized emission observed throughout the surveyed sky area are consistent with the assumptions of \cite{geil11} both in terms of intensity and RM distribution, suggesting that polarized foregrounds should not represent an insurmountable obstacle to the measurement of the EoR. In particular assuming an average polarization leakage of 2\% (Section~\ref{sec_imaging}), a Stokes $I$ leakage fainter than 20~mK would be expected in the $\alpha = 4^{\rm h}$ region, a factor of two smaller than the value simulated by \cite{geil11}.

\section{Conclusions}
\label{conc}

We have presented a 2400 square degree, Stokes $I$, $Q$, $U$ survey at 189~MHz carried out with the MWA 32 element prototype.

The survey covers $0^{\rm h} < \alpha < 6^{\rm h}$ and $21^{\rm h} < \alpha < 24^{\rm h}$, with a $20^\circ$ width in declination, centred at $\delta = -26^\circ.7$. It reaches a confusion limit of 200~mJy~beam$^{-1}$ in total intensity and a noise level of 15~mJy~beam$^{-1}$ in polarization. Our results can be summarized along the following four main themes:
\\[0.1in]

1. we exploited a novel approach to wide-field imaging with low-frequency dipole arrays that generates full-polarization images through the co-addition of snapshot images. 
As shown by \cite{mitchell12}, the integration of snapshot images is advantageous when correcting for wide-field polarization leakage in dipole arrays. In our case we achieved an average instrumental polarization better than 1.8\% over a $20^\circ$ field of view;
\\[0.1in]

2. we detected one polarized source, PMN~J0351-2744, out of a catalogue of 137 sources brighter than 4~Jy in total intensity.
Its RM~$= +34 \pm 2$~rad~m$^{-2}$ is in agreement with the value measured at 1.4~GHz. RM variations across the source cause the depolarization observed at low frequencies when integrated over the 32T PSF.
Beam depolarization also occurs for Fornax~A, for which we did not detect polarized emission at 189~MHz.

No polarized emission is detected for the remaining sources brighter than 4~Jy in total intensity. Both beam depolarization and internal Faraday dispersion could generate the observed depolarization and higher angular resolution observations will disccriminate between the two cases; 
\\[0.1in]

3. we found a wealth of patchy and filamentary diffuse polarization structures in the Galactic foreground over many degrees. The south Galactic pole and an area located at $(\ell,b) \sim (15^\circ,60^\circ)$ show the brightest polarized features, with peaks up to $\sim$13~K~RMSF$^{-1}$. There is a large sky area at $\alpha > 2^{\rm h} 30^{\rm m}$ where the polarization fluctuations are below 1~K rms.

Most of the polarized emission occurs at $0 < {\rm RM} < +10$~rad~m$^{-2}$. The limited range of RM values and the comparison with RMs of pulsars with known distances indicates that the polarization horizon is within a few hundred parsecs. 
Our data suggest that polarized emission is the result of vorticity or shear in the local ISM \citep[]{gaensler11};
\\[0.1in]

4. the polarized intensities and rotation measure distributions observed across the surveyed area are in agreement with the values simulated by \cite{geil11}, indicating that it should be possible to remove polarized foregrounds at the level required to measure the cosmological signal.

A $20^\circ \times 20^\circ$ cold patch at $\alpha = 4^{\rm h}$ looks particularly favourable because of its low polarization and because it contains the only polarized source detected, PMN~J0351-2744, which would represent an ideal polarization calibrator.

\section{Acknowledgments}
We thank an anonymous referee for helpful suggestions which improved the manuscript. This scientific work uses data obtained from the Murchison Radio-astronomy Observatory. GB thanks Matt McQuinn, Roberto Pizzo and Mario Santos for useful discussions about several topics of the present work. We acknowledge the Wajarri Yamatji people as the traditional owners of the Observatory site. Support for this work comes from the Australian Research Council (grant numbers LE0775621 and LE0882938), the National Science Foundation (grant numbers AST-0457585, AST-0821321, AST-0908884, AST-1008353 and PHY-0835713), the
U.S. Air Force Office of Scientific Research (grant number FA9550-0510247), the Australian National Collaborative
Research Infrastructure Strategy, the Australia India Strategic Research Fund, the Smithsonian Astrophysical
Observatory, the MIT School of Science, MIT Marble Astrophysics Fund, the Raman Research Institute,
the Australian National University, the iVEC Petabyte Data Store, the NVIDIA sponsored CUDA Center for
Excellence at Harvard University, and the International Centre for Radio Astronomy Research, a Joint Venture
of Curtin University of Technology and The University of Western Australia, funded by the Western Australian
State government. The Centre for All-sky Astrophysics is an Australian Research Council Centre of Excellence,
funded by grant CE110001020. The MRO is managed by the CSIRO, who also provide operational support to the MWA.

\appendix
\section{A: polarization calibration}

When an unpolarized source is observed as a calibrator, it is possible to determine the array response to unpolarized radiation by minimizing the amount of observed polarization under the assumption that it is of instrumental origin. An unknown phase rotation between the $p$ and the $q$ polarizations (in the case of linearly polarized feeds), however, can be constrained if observations of a linearly polarized calibrator are available \citep[]{sault96}. A misalingment between the $p$ and $q$ dipoles causes depolarization because linearly polarized emission is spuriously converted into circular polarization.

We used the linearly polarized source PMN~J0351-2744 to solve for the unknown delay between the $p$ and $q$ dipoles across the whole frequency bandwidth. We measured its Stokes $V$ flux through RM synthesis, under the assumption that any spurious signal has leaked from Stokes $U$ because Stokes $V$ is rotationally invariant. This is conceptually similar to the approach proposed by \cite{geil11} to remove polarized signals that leak into the EoR.

We derive the correction that minimizes the spurious Stokes $V$ flux as follows. If we define the column vector of measured Stokes parameters as ${\bf s_t}$, the vector of Stokes parameters corrected for the $p$-$q$ dipole phase difference ${\bf s_t'}$ will be:
\begin{eqnarray}
{\bf s_t'} = {\bf S^{-1}} \, {\bf M} \, {\bf S} \, {\bf s_t}
\end{eqnarray}
where ${\bf S}$ is the matrix which converts between Stokes and linear polarization coordinates \citep[see, for instance, ][]{ord10}:
\begin{eqnarray}
{\bf S} = \left [
\begin{matrix}
1 & \, \, \, \, \, 1 & \, \, \, \, 0 & \, \, \, \, 0 \\
0 & \, \, \, \, \, 0 & \, \, \, \, 1 & \, \, \, \, i \\
0 & \, \, \, \, \, 0 & \, \, \, \, 1 & \,         -i \\
1 & \,            -1 & \, \, \, \, 0 & \, \, \, \, 0 \\
\end{matrix}
\right ],
\end{eqnarray}
and ${\bf M}$ is the Mueller matrix
\begin{eqnarray}
{\bf M} = \left [ 
\begin{matrix}
1 & \, \, \, \, \, 0 & \, \, \, \, \,          0 & \, \, \, \, \,           0 \\
0 & \, \, \, \, \, e^{i \psi/2} & \, \, \, \, \, 0 & \, \, \, \, \,           0 \\
0 & \, \, \, \, \, 0 & \, \, \, \, \,          e^{-i \psi/2} & \, \, \, \, \, 0 \\
0 & \, \, \, \, \, 0 & \, \, \, \, \,          0 & \, \, \, \, \,           1 \\
\end{matrix}
\right ],
\\ \nonumber
\end{eqnarray}
where the off diagonal elements are multiplied by the phase $\psi$ which rotates the apparent Stokes $V$ signal back into Stokes $U$. The handedness of the rotation is given by the sign of the RM of PMN~J0351-2744. 

\begin{figure*}
{\epsfig{file=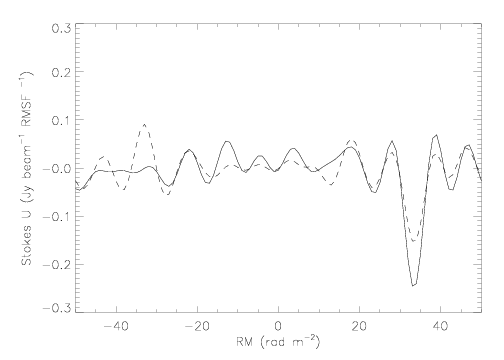, height = 7.5cm, width=8.5cm}}
{\epsfig{file=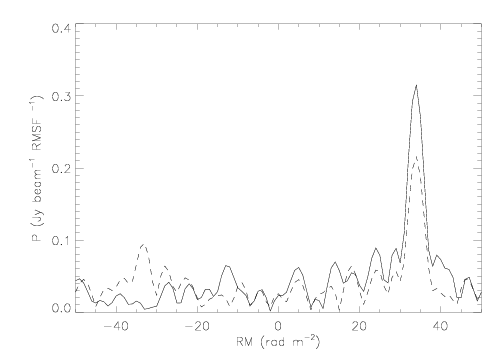, height = 7.5cm, width=8.5cm}}
\caption{Left panel: Stokes $U$ intensity for the source PMN~J0351-2744 before (dashed line) and after (solid line) the correction for the $p$-$q$ delay difference. Right panel: same as left, but for the polarized intensity $P$.\label{fig:appendix}}
\end{figure*}

The corresponding direction independent Jones matrix ${\bf J}_{DI,i}'$ for the tile $i$ used to correct the visibility data is therefore obtained by multiplying the diagonal terms of the measured ${\bf J}_{DI,i}$ by the phase terms $e^{i \psi/2}$ and $e^{-i \psi/2}$ respectively. Figure~\ref{fig:appendix} shows the polarized flux for PMN~J0351-2744 before and after the $p$-$q$ dipole phase correction.


\end{document}